\documentclass[reprint,pra,aps,notitlepage,nofootinbib]{revtex4-2}
\usepackage{amsmath}
\usepackage{amssymb}
\usepackage{bm,braket}
\usepackage{float} 
\usepackage{tikz} 
\usepackage{verbatim}
\usetikzlibrary{quantikz2}

\usepackage{hyperref}
\usepackage{multirow}
\hypersetup{
     colorlinks   = true,
     citecolor    = blue
}
\def\be{\begin{equation}}
\def\ee{\end{equation}}
\def\bea{\begin{eqnarray}}
\def\eea{\end{eqnarray}}
\def\bes{\begin{eqnarray}}
\def\ees{\end{eqnarray}}
\def\bi{\begin{itemize}}
	\def\ei{\end{itemize}} 
	
\usepackage{amsthm}

\theoremstyle{definition}

\newcommand{\vett}[1]{{\bf{#1}}}
\newcommand{\bts}{\vett{x}}
\newcommand{\groundstate}{\mathrm{G}}
\newcommand{\numberop}[1]{\hat{n}_{#1}}
\newcommand{\crt}[1]{ \hat{a}^\dagger_{#1} }
\newcommand{\dst}[1]{ \hat{a}^{\phantom{\dagger}}_{#1} }
\newcommand{\rhf}{{\mathrm{RHF}}}

\usepackage{tikz}
\usetikzlibrary{braids}

\begin{document}
\title{Efficient Quantum Chemistry Calculations on Noisy Quantum Hardware  } 

\author{Nora Bauer \thanks{Corresponding author}}
\email{nbauer@vols.utk.edu}
\affiliation{Department of Physics and Astronomy,  The University of Tennessee, Knoxville, TN 37996-1200, USA}

\author{K\"ubra Yeter-Aydeniz}
\email{kyeteraydeniz@mitre.org}
\affiliation{Quantum Information Sciences, Optics, and Imaging Department, The MITRE Corporation, 7515 Colshire Drive, McLean, Virginia 22102-7539, USA}

\author{George Siopsis}
\email{siopsis@tennessee.edu}
\affiliation{Department of Physics and Astronomy,  The University of Tennessee, Knoxville, TN 37996-1200, USA}

\date{\today}
\begin{abstract}
We present a hardware-efficient optimization scheme for quantum chemistry calculations, utilizing the Sampled Quantum Diagonalization (SQD) method. Our algorithm, optimized SQD (SQDOpt), combines the classical Davidson method technique with added multi-basis measurements to optimize a quantum Ansatz on hardware using a fixed number of measurements per optimization step. This addresses the key challenge associated with other quantum chemistry optimization protocols, namely Variational Quantum Eigensolver (VQE), which must measure in hundreds to thousands of bases to estimate energy on hardware, even for molecules with less than 20 qubits. 
Numerical results for various molecules, including hydrogen chains, water, and methane, demonstrate the efficacy of our method compared to classical and quantum variational approaches, and we confirm the performance on the \texttt{ibm-cleveland} quantum hardware, where we find instances where SQDOpt either matches or exceeds the solution quality of noiseless VQE. A runtime scaling indicates that SQDOpt on quantum hardware is competitive with classical state-of-the-art methods, with a crossover point of 1.5 seconds/iteration for the SQDOpt on quantum hardware and classically simulated VQE with the 20-qubit H$_{12}$ molecule. 
Our findings suggest that the proposed SQDOpt framework offers a scalable and robust pathway for quantum chemistry simulations on noisy intermediate-scale quantum (NISQ) devices.
\end{abstract}
\maketitle 
\twocolumngrid 
\section{Introduction}  

Quantum chemistry stands as one of the most promising applications of quantum computing, enabling accurate simulations of molecular systems that are otherwise infeasible with classical methods. The potential for quantum computers to solve problems such as electronic structure determination, reaction dynamics, and material design has inspired a surge of research, particularly in the noisy intermediate-scale quantum (NISQ) era. However, the limitations of current hardware, such as shallow circuit depths, high noise levels, and restricted qubit counts, demand innovative approaches to make quantum chemistry computations feasible.

The variational quantum eigensolver (VQE) is a hybrid quantum-classical algorithm proposed to solve Hamiltonian eigenvalue problems and first applied to quantum chemistry problems \cite{Peruzzo_2014,McClean_2016}. In most cases, this algorithm involves the preparation of a parametrized trial wavefunction, or Ansatz, on quantum hardware, and then the evaluation of the Ansatz energy. The energy is iteratively minimized in conjunction with a classical optimizer which proposes new parameter sets that are tested on quantum hardware until the minimal energy converges. VQEs have been applied to molecules such as O$_2$, CO, and CO$_2$ \cite{Sapova_2022}, but challenges in optimizing high-dimensional parameter spaces and noise resilience remain \cite{Huggins_2021}.

One of the fundamental problems in near-term quantum chemistry is the measurement budget required to measure this VQE energy, since the Hamiltonian in the Pauli basis generally consists of many non-commuting terms. Even small molecules using less than 10 qubits involve hundreds of terms \cite{Kandala_2017, nam2019groundstateenergyestimationwater}. Early studies demonstrated that molecular symmetries could be harnessed to reduce qubit requirements, enabling simulations of larger systems with fewer resources \cite{Setia_2020}. For instance, the decomposition of a 20-qubit hydrogen ring problem into smaller, solvable components highlighted the scalability of symmetry-based approaches \cite{kawashima2021optimizingelectronicstructuresimulations}. Previous approaches have also used unitary partitioning \cite{Zhao_2020}, classical shadows \cite{Preskill_2021}, overlapped grouping \cite{Wu_2023}, and molecular point group symmetries \cite{Cao_2022} to reduce the measurement budget. 
However, although these techniques can reduce the number of measurements, these approaches still struggle to bring the measurement budget for a VQE procedure to the level required for VQE on NISQ harware. 

Several foundational works have shaped the field of quantum chemistry on quantum hardware. The use of sampled quantum diagonalization (SQD) was first introduced as a method to reduce computational overhead, paving the way for efficient electronic structure simulations on limited quantum resources \cite{robledomoreno2024}. However, in these cases, the optimization of the quantum Ansatz was left to classical matrix product state (MPS) optimizers. This approach limits the path toward producing a classically intractable ground state approximation on NISQ hardware. 

Alternatively, quantum imaginary time evolution (QITE) and Quantum Lanczos (QLanczos) algorithms have been pivotal in computing ground-state and excited-state energies of molecular systems, offering practical methods for mitigating errors and improving accuracy \cite{yeteraydeniz2019practical,Tsuchimochi_2023}. These techniques, along with advancements in density-density operator-based Ansatz constructions, have been instrumental in achieving high-fidelity quantum simulations \cite{Sennane_2023}. 
Innovations in quantum machine learning (QML) for electronic structure calculations \cite{Xia_2018} and Gibbs state preparation for Rydberg atom arrays \cite{PhysRevLett.127.100504} have further expanded the methodological toolkit for quantum chemistry. Studies on benchmarking comparisons of classical spin-projected MPSs with quantum methods \cite{li2017spinprojectedmatrixproductstates} 
have demonstrated promising strategies for improving computational efficiency.

We propose a hardware-efficient framework, optimized Sampled Quantum Diagonalization (SQDOpt), that builds on the SQD method, incorporating multi-basis measurements to enhance energy estimates and optimize the quantum Ansatz on quantum hardware. By leveraging insights from prior work, including the use of optimized Ansatz states and error-resilient techniques, our approach offers a significant step forward in quantum chemistry simulations on NISQ devices. Our results indicate that the most effective near-term quantum chemistry workflow involves optimizing a variational Ansatz on quantum hardware, and then evaluating this Ansatz to obtain a high precision result \textbf{once} on classical hardware to obtain a final solution. 
We validate our methodology through extensive numerical experiments on diverse molecular systems, including hydrogen chains, water, and methane. We also test on the \texttt{ibm-cleveland} quantum hardware. 
These results highlight the efficacy of our approach in achieving high accuracy with reduced computational costs, providing a scalable and robust pathway for practical applications in quantum chemistry. 

\begin{figure}[h!]
    \centering
    \includegraphics[width=0.8\linewidth]{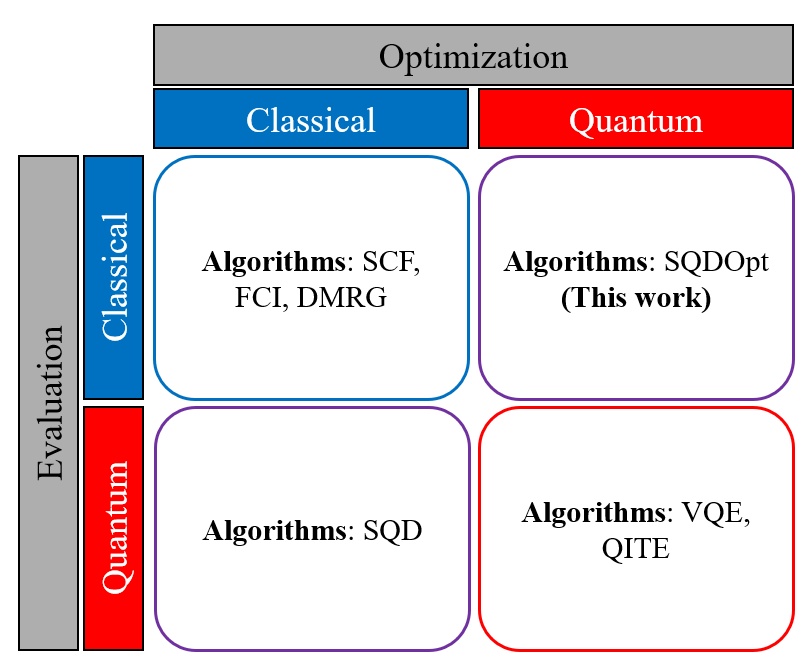}
    \caption{Comparison of SQDOpt with other commonly used algorithms/techniques for quantum chemistry calculations. }
    \label{fig:sqdidea}
\end{figure}

A comparison of our SQDOpt algorithm to the state-of-the-art algorithms and techniques, i.e., Self-consistent Field (SCF) \cite{Roothaan_1951}, Full Configuration Interaction (FCI) \cite{Ross_1952}, Density Matrix Renormalization Group (DMRG) \cite{White_1992}, VQE \cite{Peruzzo_2014,McClean_2016}, QITE \cite{McArdle_2019}, that are commonly used for quantum chemistry is presented in Fig.\ \ref{fig:sqdidea}. These algorithms were analyzed based on their use of quantum or classical resources for optimization and evaluation components of the algorithms/techniques. In this comparison, SQDOpt is the only algorithm that utilizes classical resources for evaluation and quantum resources for optimization. 

Our discussion is organized as follows. In Section \ref{sec:methods}, we provide a brief review of the SQD method and quantum Ansatz used in the calculations, and then detail how our hybrid algorithm SQDOpt builds upon this methodology by using off-diagonal measurements to better approximate energy values using fewer measurements for optimization. In Section \ref{sec:numerical}, we provide numerical results for experiments using SQDOpt, comparing to full and partial VQE and state-of-the-art classical methods. We find that in numerical simulations of 8 molecules, SQDOpt can reach lower or equal minimal energies to full VQE using only 5 measurements per optimization step in 6 of the cases, with small margins for the other 2 molecules. Furthermore, compared to classical SCF calculations, SQDOpt can provide better solutions for molecules with a higher ratio of off-diagonal terms. In Section \ref{sec:hardware}, we present results testing on the \texttt{ibm-cleveland} quantum hardware for 4 molecules, where in all cases we find instances where the SQDOpt procedure on quantum hardware produces an energy of at least comparable quality to noiseless full VQE. We also observe the scaling of the runtime for the quantum and classical methods with increasing system size, and find that SQDOpt has a crossover with classically simulated VQE at 20 qubits. Finally, in Section \ref{sec:conclusion}, we summarize our results and give direction for future research. 
\section{Methods}\label{sec:methods} 
Here, we provide a brief overview of the SQD method as proposed in \cite{robledomoreno2024}. We detail the chosen Ansatz and establish the methodology for non-diagonal SQD. 
\subsection{Sampled Quantum Diagonalization (SQD)} 
Here we outline the results of Ref.\ \cite{robledomoreno2024}. The authors chose to use the local unitary coupled Jastrow (LUCJ) Ansatz as their parametrized ansatz and optimized on classical MPS software. Then the circuit with optimal parameters was executed on a quantum computer and the output state $|\Psi\rangle$ was measured in the computational basis $N_s$ times to obtain measurement results
\begin{equation}
    \widetilde{\mathcal{X}} = \left\{ \bts \; | \; \bts \sim \widetilde{P}_\Psi (\bts) \right\}
\end{equation}
in the form of bitstrings $\bts \in \{0, 1\}^M$ distributed according to some $\widetilde{P}_\Psi$. The bitstrings represent electronic configurations (Slater determinants).

Using $K$ batches of $d$ configurations $\mathcal{S}^{(1)}\hdots , \mathcal{S}^{(K)}$ taken from the measured set $\mathcal{X}_{\textrm{R}}$, the Hamiltonian is projected and diagonalized over each $\mathcal{S}^{(k)}: k = 1, \hdots, K $. 
Each batch spans a subspace $\mathcal{S}^{(k)}$ in which the many-body Hamiltonian is projected:
\begin{equation}\label{eq:projection}
    \hat{H}_{\mathcal{S}^{(k)}} = \hat{P}_{\mathcal{S}^{(k)}} \hat{H}  \hat{P}_{\mathcal{S}^{(k)}} \textrm{, with } \hat{P}_{\mathcal{S}^{(k)}} = \sum_{\bts \in {\mathcal{S}^{(k)}}} | \bts \rangle \langle \bts | \;.
\end{equation}
The ground states and energies of $\hat{H}_{\mathcal{S}^{(k)}}$, which we label $|\psi^{(k)} \rangle$ and $E^{(k)}$, respectively, are computed using the iterative Davidson method \cite{Crouzeix_1994} .
The computational cost -- both quantum and classical -- to produce $|\psi^{(k)} \rangle$ is polynomial in $d$, the dimension of the subspace.

The ground states are then used to obtain new occupancies
\begin{equation}
\label{eq:nR_def}
    n_{p\sigma}=  \frac{1}{K}\sum_{1\leq k \leq K}  \left\langle \psi^{(k)} \right| \hat{n}_{p\sigma} \left| \psi^{(k)} \right\rangle,
\end{equation}
for each spin-orbital tuple $(p\sigma)$, averaged on the $K$ batches, where $p$ is the orbital index and $\sigma$ is the spin index. These occupancies are sent back to the configuration recovery step, and this entire self-consistent iteration is repeated until convergence, realizing a sample-based quantum diagonalization (SQD) of the target Hamiltonian. The initial guess for the $n_{p\sigma}$ values used for the first round of recovery comes from the raw quantum samples in the correct particle sector. 

On a noiseless signal $\mathcal{X}$, it is guaranteed to succeed efficiently if the ground state has a support $\mathcal{X}_\groundstate$ of polynomial size, and if the wavefunction $|\Psi\rangle$ prepared on the quantum processor has a support similar to that of the ground state. 
\subsection{Quantum Ansatz } 
In Ref.\ \cite{robledomoreno2024}, the Ansatz was optimized using MPS methods to produce a close approximation to the ground state using the local unitary coupled Jastrow (LUCJ) Ansatz 
\begin{equation}
    {| \Psi \rangle = \prod_{\mu=1}^L 
     e^{\hat{K}_\mu} e^{i \hat{J}_\mu} e^{-\hat{K}_\mu} | \bts_\rhf \rangle}.
\end{equation}
Here $\hat{K}_\mu = \sum_{pr, \sigma} K_{pr}^\mu \, \crt{p \sigma} \dst{r \sigma}$ are generic one-body operators,
$\hat{J}_\mu = \sum_{pr,\sigma\tau} J_{p\sigma, r\tau}^\mu \, \numberop{p \sigma} \numberop{r \tau}$ are density-density operators restricted to spin-orbitals that are mapped onto adjacent qubits, and $\bts_\rhf$ is the bitstring representing the restricted Hartree-Fock (RHF) state in the JW mapping. In our work, we begin with the RHF state and optimize this Ansatz over $K_{pr}^\mu$ and $J_{p\sigma, r\tau}^\mu$, starting from a random initialization, using the measured energy as the cost function.

\subsection{Beyond Diagonal Sampling: SQDOpt} 
The first approach for optimizing the Ansatz on quantum hardware would involve using the sampled Z-diagonal measurements (in the computational basis) as the cost function, which we will refer to as SQD-Z. However, this will not effectively capture the system for a Hamiltonian with large contributions from off-diagonal terms. 
The SQD-Z procedure can replace some of the optimization steps, but it would be useful if we could go further by adding more measurements besides $\langle Z_i\rangle$. An improvement would be to include the highest-weight off-diagonal measurement groups in the minimization procedure, and running the Davidson method \cite{Crouzeix_1994} calculations also in this basis. Then, a given approximation can be improved by including more measurements. This optimized SQD procedure will be referred to as SQDOpt.  

For the SQD procedure, we have control over the quantum Ansatz and the second-quantization Hamiltonian in the molecular orbital basis $\hat{H}$. It is simple to perform rotations to the quantum Ansatz to measure qubits in the $X$ and $Y$ bases. After preparing the Ansatz state $\ket{\Psi}$, the qubits to be measured in the $X$ basis have a Hadamard gate $H$ first applied, and the qubits to be measured in the $Y$ basis have a Hadamard gate, $H$, followed by an $S^\dagger$  gate applied, where 
\be H= \frac{1}{\sqrt{2}}\begin{pmatrix} 1 & 1 \\ 1 & -1 \end{pmatrix} \ , \ S = \begin{pmatrix} 1 & 0 \\ 0 & i \end{pmatrix} \ , \ S^\dagger = \begin{pmatrix} 1 & 0 \\ 0 & -i \end{pmatrix}~. \label{eq:hadamardsdagger}\ee 
Since the Hamiltonian is in the second-quantized basis, we need to convert it to the qubit basis using the Jordan-Wigner mapping, perform the rotation on the resulting qubit Hamiltonian, then perform an inverse Jordan Wigner mapping to recover a rotated second-quantized Hamiltonian. 

The second-quantized Hamiltonian is of the form 
\be \hat{\mathcal{H}}=\sum_{pr,\sigma} h_{pr}\hat{a}^\dagger_{p\sigma}\hat{a}_{r\sigma}+\sum_{prqs,\sigma\tau}\frac{(pq|rs)}{2}\hat{a}^\dagger_{p\sigma}\hat{a}^\dagger_{q\tau}\hat{a}_{s\tau}\hat{a}_{r\sigma}~, \label{eq:secondquant}\ee 
where, Roman indices ($p,r,q,s$) refer to the basis set element and Greek indices ($\sigma,\tau$) refer to the spin. 
In Jordan-Wigner encoding, the qubit mappings of the operators are given as
\begin{equation}
    \begin{split}
    &\hat{a}_{p\sigma} = \left( \prod_{q<p \ \forall \tau} Z_{q\tau}\right)\left( \prod_{\tau<\sigma} Z_{p\tau}\right) X^+_{p\sigma} ~, \\ & 
    \hat{a}_{p\sigma}^\dagger = \left( \prod_{q<p \ \forall \tau} Z_{q\tau}\right)\left( \prod_{\tau<\sigma} Z_{p\tau}\right) X^-_{p\sigma} ~,
    \end{split}
\end{equation}
where $X^\pm = \frac{1}{2}(X\pm iY)$ with $X^- = (X^+)^\dagger$. As a result of this encoding, the second-quantized fermionic Hamiltonian is mapped to 
\begin{equation}
\hat{\mathcal{H}}_{JW}=\sum_j \lambda_j h_j = \sum \lambda_j \prod_i \sigma_i^j~,
\label{eq:Hamiltonian}
\end{equation}
where $\lambda_j$ are real scalar coefficients, $h_j$ are observable tensor products of Pauli spin operators, $\sigma_i^j$ represents one of $I, X, Y$ or $Z$, $i$ indicates which qubit the operators acts on and $j$ indicates the term in the Hamiltonian.
In the Hamiltonian \eqref{eq:Hamiltonian}, we apply rotations into the appropriate basis for measurements using single-qubit unitaries. If we want to measure in a basis where $X_j$ ($Y_j$) is diagonal, then we would measure $H_j \hat{\mathcal{H}} H_j$ ($H_jS_j\hat{\mathcal{H}} S^\dagger_jH_j$), where $H_j$ is the single-qubit Hadamard gate and $S_j$, $S_j^\dagger$ are the square roots of the $Z$ gate (Eq.\ \eqref{eq:hadamardsdagger}), and $\hat{\mathcal{H}}$ is the Hamiltonian. However, since we need an operator in the second-quantization basis, we perform a reverse Jordan-Wigner mapping, 
\begin{equation}
\begin{split}
Z_{p\sigma} &= I - 2\, a^\dagger_{p\sigma} a_{p\sigma},\\[1mm]
X_{p\sigma} &= \Bigl(a^\dagger_{p\sigma} + a_{p\sigma}\Bigr)
\prod_{\substack{(q,\tau) \prec (p,\sigma)}} Z_{q\tau},\\[1mm]
Y_{p\sigma} &= i\Bigl(a^\dagger_{p\sigma} - a_{p\sigma}\Bigr)
\prod_{\substack{(q,\tau) \prec (p,\sigma)}} Z_{q\tau}\,
\end{split}
\label{eq:RJW-mapping}
\end{equation}
iteratively, until we are left with a Hamiltonian comprised of only second-quantization operators. Generally, the resulting Hamiltonian $\hat{H}_{RJW}$ will contain 1, 2, 3, and 4 operator terms. However, we will select terms of the form in \eqref{eq:secondquant} for use in the diagonalization procedure, keeping only the 2- and 4-operator terms. 

The additional bases besides the $Z$-diagonal measurement are chosen to be the highest weight groups of terms in the Hamiltonian that can be measured in a single off-diagonal basis (containing X or Y bases). On the other hand, the sparsity of the off-diagonal terms in a Hamiltonian would also impact the measurement strategy. When the off-diagonal terms are less sparse, it may be possible to measure them more efficiently by grouping them into fewer measurement bases. This is because less sparse Hamiltonian terms are more likely to commute with each other, allowing for simultaneous measurement in a single basis. In principle, partitioning measurements into non-commuting groups is NP-Hard, as it can be mapped to a graph coloring problem. We used a greedy heuristic \cite{Kosowski2008ClassicalCO} to obtain an approximate solution as implemented in the Rustworkx library \cite{Treinish2022}. 

\begin{figure} 
    \centering
    \includegraphics[width=0.99\linewidth]{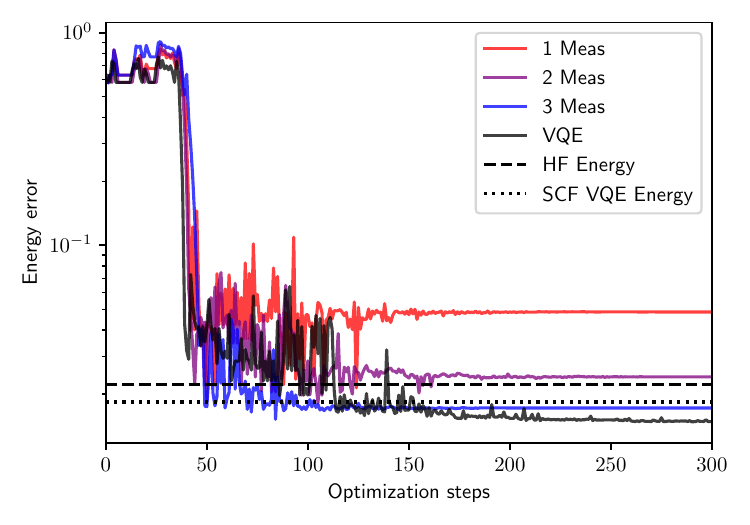}
    \caption{Results for the measured energy error (using the full Hamiltonian) of the H$_2$O molecule using the STO-3G basis set with 10 qubits. Values are shown for VQE optimization with respect to the full Hamiltonian (VQE), SQD optimization (SQDOpt) with 1 measurement (1 Meas), 2 measurements (2 Meas), 3 measurements (3 Meas), the Hartree-Fock energy (HF), and the classical FCI energy (SCF VQE). H$_2$O Hamiltonian has 156 terms, 36 noncommuting groups. } 
    \label{fig:h2o_opt_1}
\end{figure}

As an example, we consider the ground state energy of  H$_2$O molecule in Fig.\ \ref{fig:h2o_opt_1}. Here we show the error of the measured energy from the FCI energy as a function of optimization step. The optimization results for SQDOpt with 1, 2, and 3 measurements are plotted alongside a full VQE comparison. The dotted and dashed lines give the HF energy and SCF-optimized VQE energy, respectively. Using only 1 measurement per optimization step (red), the converged energy error is higher than even the classical HF energy. The 2-measurement result (purple) gives an improvement, and the 3-measurement result (blue) outperforms both the HF and SCF VQE energies. The full VQE (black) obtains a slightly lower energy; however, the full VQE simulation requires measurements in 36 noncommuting bases per step, as opposed to 3 measurements required for SQDOpt. 

\section{Numerical Results } \label{sec:numerical}

Here we show results for various molecules using the SQDOpt approach, starting with Hydrogen chains. The optimization procedure uses the constrained optimization by linear approximation (COBYLA) \cite{Powell1994ADS} minimization protocol with a maximum of 500 optimization steps. For the following results, the Hydrogen chains have the 2 lowest orbitals frozen. The SQD optimization uses 5 measurements per iteration, chosen by sorting the Pauli string terms of the Hamiltonian into groups that can be measured in a single basis, then selecting the 5 highest weight bases. In all cases studied, this included the Z-diagonal basis. 

For comparison, the FCI energy is computed, which is the exact minimal energy for the choice of basis and active space for a given molecule. Additionally, the VQE algorithm consisting of measurements in all bases is run, in addition to the partial VQE algorithm which minimizes using only measurements in 5 chosen bases. SQD-Z was also run to obtain an energy using only the single Z-diagonal measurement. Finally, the SCF calculation was run to optimize the VQE Ansatz (SCFVQE). The FCI and SCF calculations were performed with the PySCF package \cite{PYSCF}, and the VQE and partial VQE were performed using the FFSim package \cite{ffsim}. 

\begin{figure*}[ht!]
    \centering
    \includegraphics[width=0.8\linewidth]{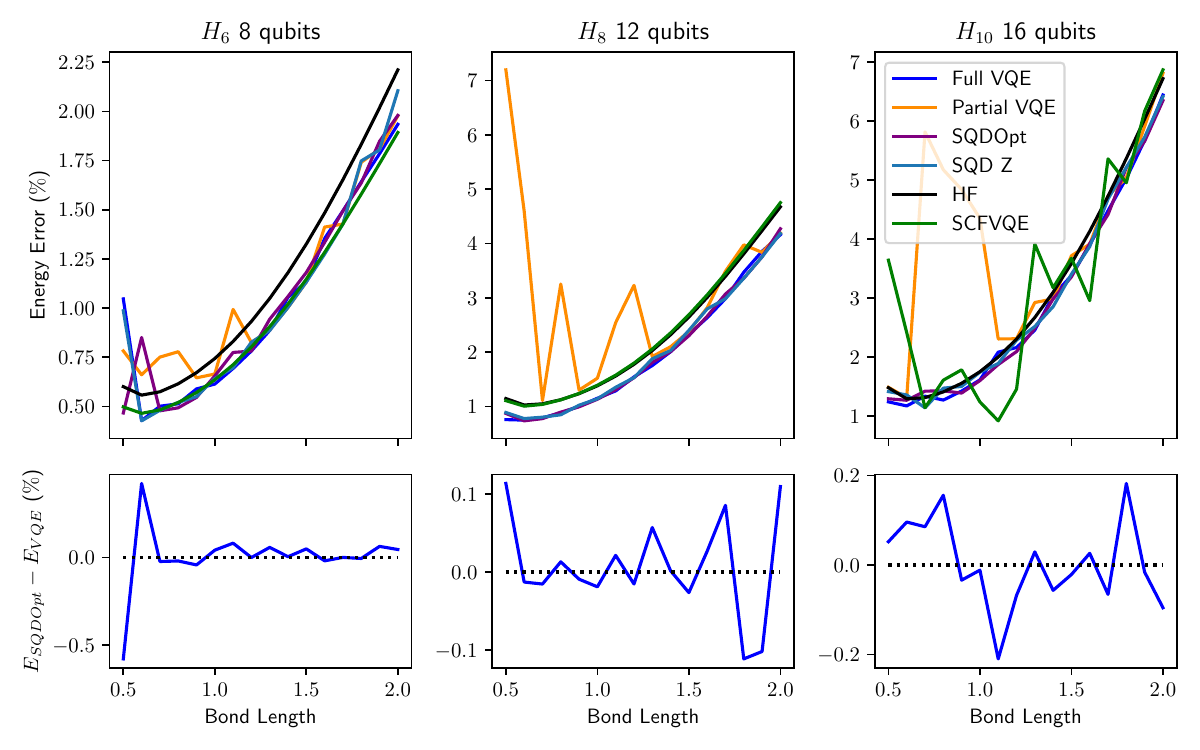}
    \caption{Upper panel: Results for the measured energy (using the full Hamiltonian) of the H$_6$, H$_8$, and H$_{10}$ molecules, respectively, using the STO-3G basis set with 2 frozen orbitals (FO). Energies are shown for VQE optimization with respect to the full Hamiltonian (VQE), SQD optimization (SQDOpt), the Hartree-Fock energy (HF), and the classical SCF VQE solution energy (SCF VQE). Lower panel: Energy differences between the SQDOpt and VQE energies expressed as a percentage of the FCI energy as a function of the bond length. } 
    \label{fig:hc}
\end{figure*}

The energy percent error from the FCI energy using SQDOpt (purple) is shown in Fig.\ \ref{fig:hc} as a function of bond length for H$_6$, H$_8$, and H$_{10}$. These values are compared with the full VQE result (blue), the partial VQE (orange), and SQD-Z (cyan). All of these results are compared with the HF energy (black) and SCF-computed VQE energy (green). The lower subplots show the difference between the energies obtained from the SQDOpt procedure and the full VQE procedure. 

In these cases, the SQDOpt results are comparable to the full VQE results, with an average slightly lower energy for each hydrogen chain. The partial VQE results have much higher error than even the HF state, especially in the small bond length regime. The SQD Z errors are on average higher than the SQDOpt error values. 

Next, we consider the ground state energies of N$_2$ and O$_2$ dimer molecules, as well as H$_2$O and CH$_4$, all in the STO-3G basis set with 2 frozen orbitals. A summary of all numerical results at the approximate equilibrium bond distances is given in Table \ref{tab:tab1}. In 6 of the 8 instances, SQDOpt produces a lower or equal energy error to full VQE, with the other 2 instances divided by a small margin. In all cases, SQDOpt produced a lower error than the classically-optimized SCF VQE calculation. 

\begin{figure}
    \centering
    \includegraphics[width=1\linewidth]{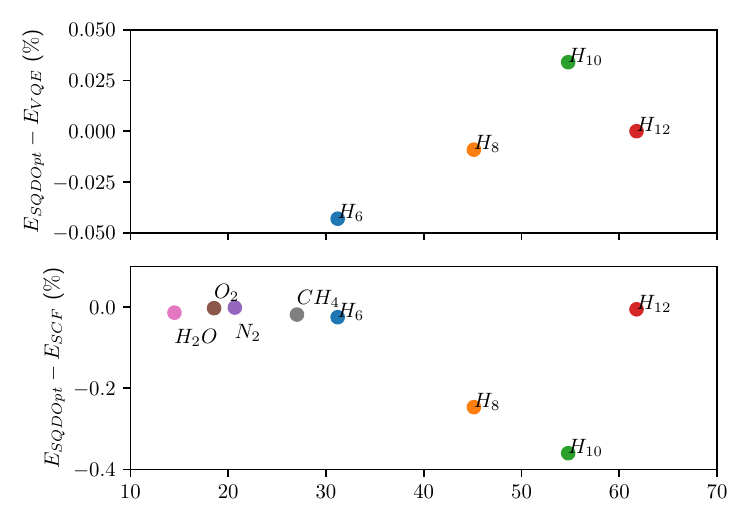}
    \caption{Upper panel: Difference between the SQDOpt energy ($E_{\mathrm{SQDOpt}}$) and the VQE energy ($E_{\mathrm{VQE}}$) as a function of the percentage of off-diagonal Hamiltonian terms (by magnitude) for the Hydrogen chains studied at the equilibrium bond distance. Lower panel: Difference between the SQDOpt energy ($E_{\mathrm{SQDOpt}}$) and the SCF energy ($E_{\mathrm{SCF}}$) as a function of the percentage of off-diagonal Hamiltonian terms (by magnitude) for the molecules studied. } 
    \label{fig:hc_4}
\end{figure}

An interesting metric is the ratio of the magnitude of off-diagonal terms in the Hamiltonian to the full set of terms (omitting identity terms). The difference between the SQDOpt energy ($E_{\mathrm{SQDOpt}}$) and the VQE energy ($E_{\mathrm{VQE}}$) as a function of the percentage of off diagonal terms in the Hamiltonian is given in Fig.\ \ref{fig:hc_4}. For these more difficult problems with larger off-diagonal contributions, the SQDOpt result and the VQE result are very close.

\begin{table*}
    \centering
    \begin{tabular}{|c||c|c|c|c|c|c|c|c|c|}\hline\hline 
    \textbf{Molecule}  &  $H_6$ &  $H_8$ &  $H_{10}$ &  $H_{12}$ & $N_2$  &  $O_2$  &  $H_2O$  &  $CH_4$ \\ \hline\hline 
Bond length & 0.9 & 0.9 & 0.9 & 0.9 & 1.2 & 1.3 & 1.9 & 1 \\ \hline 
VQE Err. (\%)  &  0.589 &  1.0021 &  \textbf{\textcolor{blue}{1.3933}} & 1.7834 & \textbf{\textcolor{blue}{0.1445}} &  0.1263 &  0.0313 &  0.2969 \\ \hline 
Partial VQE Err. (\%) &  0.6453 &  1.3011 &  4.84 & 8.6664 & 0.1762 &  0.2274 &  0.1347 &  0.3583 \\ \hline 
SQDOpt Err. (\%) &  \textbf{\textcolor{blue}{0.546}} &  \textbf{\textcolor{blue}{0.9929}} &  1.4273 & \textbf{\textcolor{blue}{1.7834}} & 0.1489 &  \textbf{\textcolor{blue}{0.1211}} &  \textbf{\textcolor{blue}{0.0313}} &  \textbf{\textcolor{blue}{0.2968}} \\ \hline 
SQD Z  Err. (\%) &  0.5496 &  1.0235 &  1.5055 & 1.7834 & 0.1523 &  0.1261 &  0.0314 &  0.2977 \\ \hline 
HF Err. (\%) &  0.6718 &  1.2352 &  1.5597 & 1.7576 & 0.1758 &  0.1464 &  0.0534 &  0.3258 \\ \hline 
SCF Err. (\%) &  0.571 &  1.2395 &  1.7872 & 1.7889 & 0.1502 &  0.1238 &  0.0451 &  0.3154 \\ \hline 
SQD-VQE (\%) &  -0.0431 &  -0.0091 &  0.034 & 0 & 0.0044 &  -0.0052 &  -1.2E-5 &  -0.0001 \\ \hline 
SQD-SCF (\%) &  -0.025 &  -0.2466 &  -0.3599 & -0.0055 & -0.0013 &  -0.0027 &  -0.0138 &  -0.0185 \\ \hline 
\end{tabular}
    
    \caption{Summary of results for various molecules, given as the percent errors at the approximate equilibrium bond length. The lowest percent errors between VQE and SQDOpt are given in blue and bold text.  } 
    \label{tab:tab1}
\end{table*}

We can also look at this performance of these molecules compared to the classical SCF energy as the problem difficulty increases. The difference between the SQDOpt energy ($E_{\mathrm{SQDOpt}}$) and the SCF energy ($E_{\mathrm{SCF}}$) as a function of the percentage of off diagonal terms in the Hamiltonian is given in Fig.\ \ref{fig:hc_4}. For these more difficult problems with larger off-diagonal contributions, SDQOpt has superior performance to the classical SCF procedure. 

To test the performance of SQDOpt on NISQ hardware, we ran simulations on both the noisy simulator and \texttt{ibm-cleveland} quantum hardware, discussed in the next Section. 


\section{Hardware Results}\label{sec:hardware} 

Here, we discuss testing of our procedure on the \texttt{ibm-cleveland} quantum hardware. 
We tested both the quality of the measured ground state energy given an optimized set of parameters and the quality of a set of parameters when optimized on hardware. 

\begin{figure}[ht!]
    \centering
    \includegraphics[width=\linewidth]{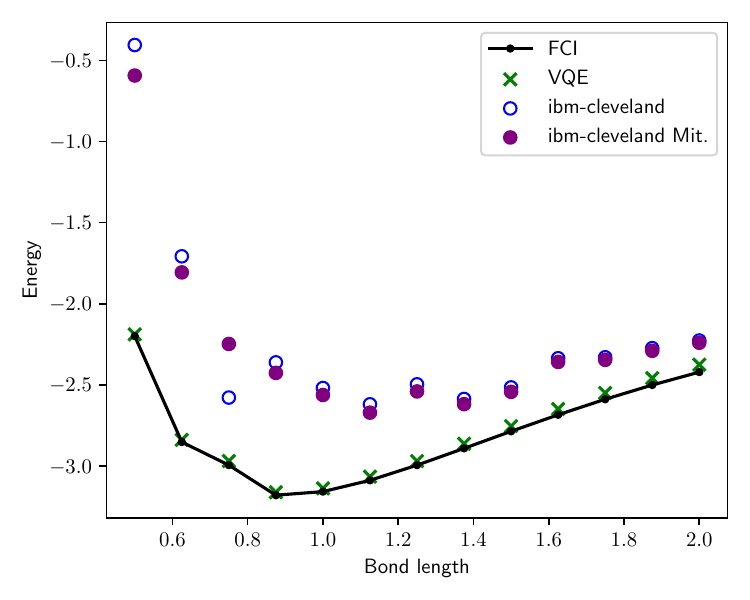}
    \caption{Quantum hardware results on \texttt{ibm-cleveland} results for ground state energy of the $H_6$ molecule using STO-3G basis and 2 frozen orbitals, which has 68 noncommuting groups. The raw (blue circles) and measurement error-mitigated (purple circles) results from \texttt{ibm-cleveland} are compared with the SCF-optimized VQE energy (green $\times$'s) (VQE) and the FCI energy (black dots).  } 
    \label{fig:h6_cleveland_meas1}
\end{figure}

First, we measured the ground state energy of a molecule using the full number of measurements. Results for measuring the energy of the H$_6$ molecule on \texttt{ibm-cleveland} are shown in Fig.\ \ref{fig:h6_cleveland_meas1}. This experiment used the optimal parameters (as can be found with SQDOpt) and the complete measurement basis, which, for H$_6$, consists of 68 non-commuting groups of operators. There is a significant amount of error in the energy, especially when the bond length is small, indicating that evaluation of the energy is still quite far away from obtaining chemical accuracy on NISQ hardware. If measurement error mitigation \cite{m3} is included, there is a slight improvement in the energy estimates, so the error source is likely from gate error or coherence time. Note that in the measurement mitigation used here, the confusion matrix used in the mitigation was computed once before the experiments, not at each step, which could contribute to the lack of significant difference between the raw and mitigated values. 

\begin{figure}[ht!]
    \centering
    \includegraphics[width=\linewidth]{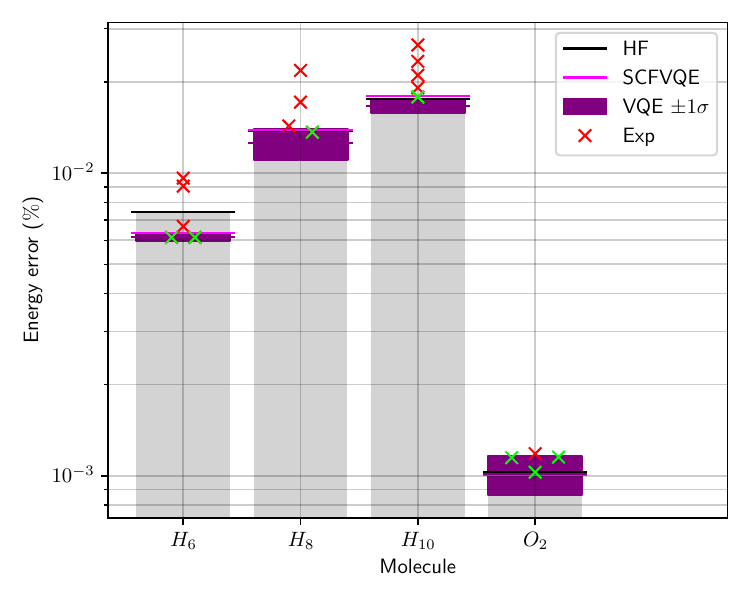}
    \caption{Quantum hardware results on \texttt{ibm-cleveland} for ground state energy optimization of molecules H$_6$, H$_8$, H$_{10}$, O$_2$ using SQDOpt. The experimental (Exp.) results from \texttt{ibm-cleveland} (red/lime $\times$'s) are compared with the HF energy (black), the SCF VQE energy (magenta), and the full VQE energy with 1 standard deviation (purple). Here, the lime $\times$'s indicate a solution at least as good as the VQE energy range, while red $\times$'s indicate that the solution was not within the range. } 
    \label{fig:h6_cleveland_meas2}
\end{figure}

However, the main objective of our procedure is for the optimization of quantum Ans\"atze on quantum hardware. Thus, we turn to running the SQDOpt procedure on hardware. We again use the COBYLA optimizer with maximum number of 500 optimization steps, where each SQDOpt step involves measurements in the top 5 bases. The results for the H$_6$, H$_8$, H$_{10}$, and O$_2$ molecules are given in Fig.\ \ref{fig:h6_cleveland_meas2}. Here we compare the SQDOpt ($\times$'s) results on the hardware to the HF energy, the SCF-optimized VQE parameter energy (SCFVQE), and the noiseless VQE energy obtained using the same number of optimization steps, but with the full measurement set. The lime $\times$'s indicate that the solution was within the VQE range, and the red $\times$'s indicate solutions outside this range. 

For all molecules studied, there was at least one case where the SQDOpt solution \textbf{optimized on the quantum hardware} found a solution at least as good as the VQE range (indicated by the lime $\times$'s on Fig.~\ref{fig:h6_cleveland_meas2}). For H$_6$, there were two of the five instances where the SQDOpt result on quantum hardware found a lower energy than the noiseless VQE simulations. 

\begin{figure}[ht!]
    \centering
    \includegraphics[width=\linewidth]{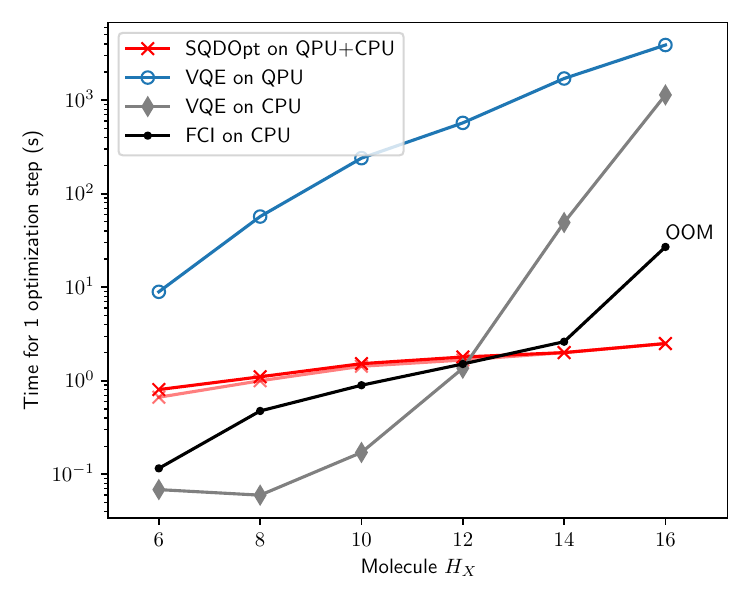}
    \caption{Comparison of the runtime per optimization step for Hydrogen chain molecules using our hybrid algorithm SQDOpt on \texttt{ibm-cleveland} (red) and VQE on \texttt{ibm-cleveland} (blue) and a 24-core Intel Gold 6248R CPU with 192 GB RAM (grey). The FCI runtime is shown in gray, although this runtime is for the entire procedure, not 1 iteration. The faded red line below the SQDOpt result indicates the runtime of only the quantum hardware step, omitting the Davidson iteration on classical hardware. } 
    \label{fig:sqdopt_benchmark}
\end{figure}

Finally, we studied the scaling of the quantum and classical algorithms with increasing system sizes. Fig.\ \ref{fig:sqdopt_benchmark} compares the runtime per optimization step of the SQDOpt procedure using \texttt{ibm-cleveland} quantum hardware and a CPU (red) with VQE on \texttt{ibm-cleveland} quantum hardware (blue) and simulated VQE on a CPU node (grey) using the FFSim software \cite{ffsim}. The classical CPU was a 24-core Intel Gold 428R CPU node with 192 GB RAM. The quantum estimates were calculated from IBM QPU and include the time for 1000 shots, without queue time. For SQDOpt, the time includes the quantum runtime for the sampling on the QPU and the runtime of the Davidson diagonalization procedure on the classical node. The faded red line below the SQDOpt line indicates the runtime for just the QPU tasks, which take the majority of the runtime. The total runtime for the FCI calculation on the same classical node (black) is also included. The FCI calculation encountered an out-of-memory (OOM) error for molecules larger than H$_{16}$. Our SQDOpt procedure had a modest scaling with system size, as the number of measurements per step was constant. Since VQE required more basis measurements as the system size increased, it had a poor scaling with system size. Simulated VQE runtime grew quickly with system size and had a runtime crossover with SQDOpt at H$_{12}$ (20 qubits). Since FCI is an exact calculation and the given time is for the entire procedure (not just 1 optimization step), this is the best method up to H$_{16}$, after which the memory requirements are exceeded and larger CPUs are required. Therefore, after these memory requirements are exceeded on large molecules, we expect SQDOpt to provide the most efficient runtime scaling compared to VQE on QPU or CPU. 

\section{Conclusion} \label{sec:conclusion}

In this work, we presented a hardware-efficient framework for quantum chemistry ground state calculations based on the SQD method. By combining classical-quantum feedback loops with multi-basis measurements, we demonstrated a scalable approach to significantly reduce the number of measurements required for accurate molecular energy computations. We tested numerically on molecules up to 20 qubits and found that our hybrid algorithm SQDOpt matched or exceeded the quality of the standard VQE algorithm and consistently performed better than the SCF VQE classical heuristic algorithm. On quantum hardware (\texttt{ibm-cleveland}), we found instances in all 4 molecules tested where SQDOpt matched noiseless VQE solution quality, in some cases exceeding it. Finally, we performed runtime scaling and found evidence of a crossover where SQDOpt on hardware takes less time than simulated VQE on a cluster node. 

Our main priority moving forward is pushing the SQDOpt algorithm to the limit where exact classical calculations require too much memory, and we have the potential for runtime quantum advantage. Additionally, it is imperative to quantify the scaling of the number of basis measurements per SQDOpt step to maintain the quality of solution as system size increases. For these larger molecules, a fair runtime comparison would also involve considering parallelization and GPU resources in the classical calculations, since these tools offer runtime advantages. 

As quantum hardware continues to evolve, the integration of fault-tolerant quantum algorithms into the SQD framework could further enhance its scalability and accuracy, particularly for larger and more complex molecular systems. Moreover, the application of multi-basis measurements and advanced sampling techniques could be extended to address other challenges in quantum chemistry, such as reaction dynamics \cite{Kale_2024} and excited-state properties \cite{Cadi_Tazi_2024}. Collaboration with industrial partners could explore the potential of this methodology for energy-efficient materials design and drug discovery.

In addition, it is worth focusing on improving the efficiency of classical components, such as the optimization routines used in the iterative feedback loop. Developing hybrid algorithms that synergize SQD with machine learning techniques may offer new strategies for parameter optimization and error correction. Furthermore, extending the approach to incorporate real-time dynamics simulations and open quantum systems could significantly broaden its applicability.

As quantum computing progresses toward fault-tolerant devices, the lessons learned from hardware-efficient approaches will remain valuable. By bridging the gap between theoretical advancements and hardware limitations, our work provides a foundational step toward realizing the full potential of quantum chemistry in the NISQ era and beyond.

\acknowledgments

This material is based upon work supported by the U.S. Department of Energy, Office of Science, Office of Science, Advanced Scientific Computing Research (ASCR) program, under Awards DE-SC0024325 and DE-SC0024451. We also
acknowledge support by the Cleveland Clinic Foundation, and the U.S. National Science Foundation under
Award DGE-2152168. A
portion of the computation for this work was performed on the University of Tennessee Infrastructure for Scientific Applications and Advanced Computing (ISAAC)
computational resources. This
research also used resources of the Oak Ridge Leadership Computing Facility, which is a DOE Office of Science User Facility
supported under Contract DE-AC05-00OR22725.

©2025 The MITRE Corporation. ALL RIGHTS RESERVED. Approved for public release. Distribution unlimited. PR$-$25$-$0695.

%

\end{document}